\documentstyle[aps,epsfig]{revtex}
\begin{document}
%\draft

\title{Role of effective interaction in nuclear disintegration processes }

\author{D.N. Basu\thanks{E-mail:dnb@veccal.ernet.in}}
\address{Variable  Energy  Cyclotron  Centre,  1/AF Bidhan Nagar,
Kolkata 700 064, India}
\date{\today }
\maketitle
\begin{abstract}

Abstract: A simple superasymmetric fission model using microscopically calculated nuclear potentials has shown itself to be outstandingly successful in describing highly asymmetric spontaneous disintegration of nuclei into two composite nuclear fragments. The nuclear interaction potentials required to describe these nuclear decay processes have been calculated by double folding the density distribution functions of the two fragments with a realistic effective interaction. The microscopic nucleus-nucleus potential thus obtained, along with the Coulomb interaction potential and the minimum centrifugal barrier required for the spin-parity conservation, has been used successfully for the lifetime calculations of these nuclear disintegration processes.

 

\end{abstract}


%\eject

      The highly asymmetric spontaneous disintegration of nuclei into two composite nuclear fragments has intrigued physicists for many years. Such nuclear disintegration processes known as cluster decays include, first of all but not exclusively, the $\alpha$-decay. First experimental observation of $\alpha$ radioactivity \cite{r1} was made several decades ago. A theoretical explanation for $\alpha$ radioactivity in terms of quantum mechanical barrier penetration \cite{r2} had been, at least, qualitatively successful. There has recently been a renewed interest in the cluster decays primarily motivated by an increase in the role of $\alpha$-decay in the spectroscopy of the unstable nuclei \cite{r3} and by the discovery of the exotic radioactivity \cite{r4} such as spontaneous emission of heavier clusters like $^{14}C,~ ^{20}O,~ ^{14}Ne$, etc. Predictions for $\alpha$ and various exotic decays have been made by the analytical superasymmetric fission model (ASAFM) \cite{r5,r6} with reasonable success. This was followed by the preformed cluster model (CM) calculations for the $\alpha$ decay \cite{r7} with similar success. But  both the theoretical approaches described above use rather unusual forms of phenomenological potentials for the nucleus-nucleus interactions. The ASAFM uses a parabolic potential approximation for the nuclear interaction within the superasymmetric fission model (SAFM) description which yields analytical expressions for the decay lifetimes, while the CM uses a cos-hyperbolic form of nuclear interaction potential  \cite{r7} designed to fit the experimental data. 

       In the CM the cluster is assumed to be formed before it penetrates the barrier and its preformation probability is also included in the calculations. Though the physics of the CM and the SAFM descriptions are apparently different, but in fact they are very similar. Interpreting the cluster preformation probability within a fission model as the penetrability of the pre-scission part of the barrier, it was shown that the cluster model is equivalent to the fission model \cite{r8}. In the present work, the nuclear potentials needed for the SAFM have been obtained microscopically by folding in the nuclear density distributions of the two composite nuclear fragments with the realistic M3Y effective interaction. Any liquid drop like properties such as surface energy are basically macroscopic manifestation of microscopic phenomena. A double folding potential obtained using M3Y effective interaction is more appropriate because of its microscopic nature. A potential energy surface is inherently embedded in this description. The minimum centrifugal barrier that has been used for the lifetime calculations has been fixed by the requirement of the spin-parity conservation. This simple SAFM using microscopically calculated nuclear potentials has been found to provide excellent estimates for the half lives of $\alpha$ and various heavier cluster decays.  

      The microscopic nuclear potentials $V_N(R)$ have been obtained by double folding in the densities of the emitted cluster and the residual daughter nucleus with the finite range realistic M3Y effective interacion as

\begin{equation}
 V_N(R) = \int \int \rho_1(\vec{r_1}) \rho_2(\vec{r_2}) v[|\vec{r_2} - \vec{r_1} + \vec{R}|] d^3r_1 d^3r_2 
\label{seqn1}
\end{equation}
\noindent
where $\rho_1$ and $\rho_2$ are the density distribution functions for the two composite nuclear fragments. The density distribution function in case of $\alpha$ particle has the Gaussian form

\begin{equation}
 \rho(r) = 0.4229 exp( - 0.7024 r^2)
\label{seqn2}
\end{equation}                                                                                                                                           \noindent     
whose volume integral is equal to $A_\alpha ( = 4 )$, the mass number of $\alpha$-particle. The density distribution function in case of heavier fragment has been chosen to be of the spherically symmetric form given by

\begin{equation}
 \rho(r) = \rho_0 / [ 1 + exp( (r-c) / a ) ]
\label{seqn3}
\end{equation}                                                                                                                                           \noindent     
where                        
 
\begin{equation}
 c = r_\rho ( 1 - \pi^2 a^2 / 3 r_\rho^2 ), ~~    r_\rho = 1.13 A_d^{1/3}  ~~   and ~~    a = 0.54 ~ fm
\label{seqn4}
\end{equation}
\noindent
and the value of $\rho_0$ has been fixed by equating the volume integral of the density distribution function to the mass number of the fragment. The finite range M3Y effective interaction $v(s)$ appearing in the eqn.(1) is given by \cite{r9} 

\begin{equation}
 v(s) = 7999 \exp( - 4s) / (4s) - 2134 \exp( - 2.5s) / (2.5s)
\label{seqn5}
\end{equation}   
\noindent
This interaction is based upon a realistic G-matrix. Since the G-matrix was constructed in an oscillator representation, it is effectively an average over a range of nuclear densities and therefore the M3Y has no explicit density dependence. For the same reason there is also an average over energy and the M3Y has no explicit energy dependence either. The only energy dependent effects that arises from its use is a rather weak one contained in an approximate treatment of single-nucleon knock-on exchange. The success of the extensive analysis indicates that these two averages are adequate for the real part of the optical potential for heavy ions at energies per nucleon of $< 20MeV$. However, it is important to consider the density and energy dependence explicitly for the analysis of $\alpha$-particle scattering at higher energies ($>100 MeV$) where the effects of a nuclear rainbow are seen and hence the scattering becomes sensitive to the potential at small radii. Such cases were studied introducing suitable and semirealistic explicit density dependence \cite{r10,r11} into the M3Y interaction which was then called the DDM3Y and was very successful for interpreting consistently the high energy elastic $\alpha$ scattering data. Since the released energies involved in the cluster decay processes are very small compared to the energies involved in high energy heavy ion scattering, these effects are expected to be small for processes like cluster radioactivity. The total interaction energy $E(R)$ between the emitted nucleus and the residual daughter nucleus is equal to the sum of the nuclear interaction energy, the Coulomb interaction energy and the centrifugal barrier. Thus

\begin{equation}
 E(R) = V_N(R) + V_C(R) + \hbar^2 l(l+1) / (2\mu R^2)
\label{seqn6}
\end{equation}   
\noindent
where $\mu = mA_e A_d/A$  is the reduced mass, $A_e, A_d, A$ are the mass numbers of the emitted cluster, residual daughter nucleus and the parent nucleus respectively and m is the nucleon mass measured in the units of $MeV/c^2$. The minimum angular momentum $l_{min}$ carried away by the emitted particle is decided by the requirement of the spin-parity conservation which in turn decides the minimum centrifugal barrier that has been used for the lifetime calculations of these nuclear disintegation processes. Assuming spherical charge distribution for the residual daughter nucleus and considering the emitted cluster to be a point charge, the Coulomb potential $V_C(R)$ has been taken as 

\begin{eqnarray}
 V_C(R) =&&Z_e Z_d e^2/ R~~~~~~~~~~~~~~~~~~~~~~~~~~~~~~~~~~for~~~~R \geq R_c \nonumber\\
            =&&(Z_e Z_d e^2/ 2R_c).[ 3 - (R/R_c)^2]~~~~~~~~~~for~~~~R\leq R_c 
\label{seqn7}
\end{eqnarray}   
\noindent
where $Z_e$ and $Z_d$ are the atomic numbers of the emitted-cluster and the daughter nucleus respectively. The touching radial separation $R_c$ between the emitted-cluster and the daughter nucleus is given by $R_c = c_e+c_d$ where $c_e$ and $c_d$ have been obtained using eqn.(4). The energetics allow spontaneous emission of clusters only if the released energy 

\begin{equation}
 Q = M - ( M_e + M_d)
\label{seqn8}
\end{equation}
\noindent
is a positive quantity, where $M$, $M_e$ and $M_d$ are the atomic masses of the parent nucleus, the emitted cluster and the residual daughter nucleus, respectively,  expressed in the units of energy. It is important to mention here that the correctness of predictions for possible decay modes, therefore, rests on the accuracy of ground state masses of nuclei.

      In the present model (SAFM), the half life of the parent nucleus against the split into an emitted cluster and a residual daughter nucleus has been calculated using the WKB barrier penetration probability. The zero point vibration energy which is a quantum mechanical phenomena arising out of the finite size of the nucleus represents non zero ground state energy of the quantum oscillator. For a quantum oscillator (consisting of the daughter nucleus and the cluster nucleus to be emitted) the zero point vibration  energy is directly related to the assault frequency. The assault frequency $\nu$ has, therefore, been obtained from the zero point vibration energy using the relationship $E_v = (1/2)\hbar\omega$ where $\omega=2\pi\nu$. The half life $T$ of the parent $(A, Z)$ nucleus against its split into an emitted $(A_e, Z_e)$ cluster and a daughter $(A_d, Z_d)$ nucleus is given by

\begin{equation}
 T = [(h \ln2) / (2 E_v)] [1 + \exp(K)]
\label{seqn9}
\end{equation}
\noindent
where the action integral $K$ within the WKB approximation is given by \cite{r6}

\begin{equation}
 K = (2/\hbar) \int_{R_a}^{R_b} {[2\mu (E(R) - E_v - Q)]}^{1/2} dR
\label{seqn10}
\end{equation}
\noindent
where $R_a$ and $R_b$ are the two turning points of the WKB action integral determined from the equations

\begin{equation}
 E(R_a)  = Q + E_v =  E(R_b)
\label{seqn11}
\end{equation} 

      The zero point vibration energies used in the present calculations are the same as that described in reference \cite{r12} immediately after eqn.(4) for the $\alpha$ cluster and by eqns.(5) for the heavier clusters. The shell effects for every cluster radioactivity are implicitly contained in the zero point vibration energy due to its proportionality with the Q value, which is maximum when the daughter nucleus has a magic number of neutrons and protons. Values of the proportionality constants of $E_v$ with $Q$ is the largest for even-even parent and the smallest for the odd-odd one. Other conditions remaining same one may observe that with greater value of $E_v$, lifetime is shortened indicating higher emission rate. The two turning points of the action integral given by eqn.(10) have been obtained by solving eqns.(11) using the microscopic double folding potential given by eqn.(1) along with the Coulomb potential and the minimum centrifugal barrier determined from the spin parity conservation. Then the WKB action integral between these two turning points has been evaluated numerically using eqn.(1), eqn.(6), eqn.(7) and eqn.(8). Finally, the half lives of the cluster decays have been calculated using eqn.(9).  

      For present calculations entire sets of experimental data for the $\alpha$ decay half lives of references \cite{r13}, \cite{r14} and  \cite{r6}, respectively, have been chosen for comparison with the present theoretical calculations for which the experimental ground state masses for the parent and daughter nuclei are available. Older experimental values have been substituted by the recent ones \cite{r15} and the spins-parities of the parent and daughter nuclei along with the minimum angular momenta $l_{min}$ carried away by the $\alpha$-particles calculated considering spin-parity conservation have been listed. The uncertain assignments of spins-parities have been shown within parentheses and unknown values have been left blank. Experimental values for $\alpha$ decay half lives have been presented in Table 1 along with corresponding results of the present SAFM calculations with microscopic potentials calculated using the M3Y effective interaction. Results of the present calculations using the density dependent M3Y effective interaction (DDM3Y) supplemented by a zero-range pseudo potential have been shown inside parentheses. In DDM3Y the effective nucleon-nucleon interaction $v(s)$ is assumed to be density and energy dependent and therefore becomes functions of density and energy and is given by 

\begin{equation}
  v(s,\rho_1,\rho_2,E) = t^{M3Y}(s,E)g(\rho_1,\rho_2,E)
\label{seqn12}
\end{equation}   
\noindent
where $t^{M3Y}$ is the same M3Y interaction given by eqn.(5) but supplemented by a zero range pseudo-potential \cite{r10}

\begin{equation}
  t^{M3Y} = 7999 \exp( - 4s) / (4s) - 2134 \exp( - 2.5s) / (2.5s) + J_{00}(E) \delta(s)
\label{seqn13}
\end{equation}   
\noindent
where the zero-range pseudo-potential representing the single-nucleon exchange term is given by

\begin{equation}
 J_{00}(E) = -276 (1 - 0.005E / A_\alpha ) (MeV.fm^3)
\label{seqn14}
\end{equation}   
\noindent
and the density dependent part has been taken to be \cite{r11}

\begin{equation}
 g(\rho_1, \rho_2, E) = C (1 - \beta(E)\rho_1^{2/3}) (1 - \beta(E)\rho_2^{2/3})
\label{seqn15}
\end{equation}   
\noindent
which takes care of the higher order exchange effects and the Pauli blocking effects. The energy E appearing in the above equations is the energy measured in the centre of mass of the emitted cluster - daughter nucleus system and for the cluster decay process it is equal to the released energy Q. Since the released energies involved in the cluster decay processes are very small compared to the energies involved in high energy heavy ion scattering, the $\beta(E)$ has been considered as a constant and independent of energy and has been found to be equal to 1.6 obtained from optimum fit to the data. The zero-range pseudo-potential $J_{00}(E)$ is also practically independent of energy for the cluster decay processes and has be taken as $-276 MeV.fm^3$. Results of calculations of ASAFM(1986), ASAFM(1991), the Viola-Seaborg parametrization with Sobiczewski et.al.constants (VSS) \cite{r16} and the liquid drop model (LDM)  \cite{r17} have also been presented. The ASAFM(1986), ASAFM(1991) and the VSS have been recalculated with the exact Q values listed in the Table 1. The chi-squares per degrees of freedom ($\chi^2/F$) have been calculated assuming uniform percentage of experimental error which guarantees equal weights for all the data.   
 
\begin{table}
\caption{Comparison between Measured and Calculated $\alpha$ decay Half-Lives}
\begin{tabular}{cccccccccccc}
Parent & Parent &Parent&Daughter&ASAFM(86)&ASAFM(91)&VSS(89)&LDM(01)&Present&Expt.&Energy&$l_{min}$  \\ 
 & & & .& & & & &M3Y (DDM3Y) & &Released &\\ \hline
Z&A&$J^{\pi}$&$J^{\pi}$&$log_{10}T(s)$&$log_{10}T(s)$&$log_{10}T(s)$&$log_{10}T(s)$&$log_{10}T(s)$&$log_{10}T(s)$&Q(MeV)& \\ \hline

  87& 221&5/2- &9/2-& 2.78 &    2.73 &    2.76 &2.20&    2.33 (2.43)&    2.47$^a$&       6.47 &  2 \\
  88& 221& 5/2+&9/2+&1.55 &    1.67   &  1.82   & 1.53& 1.27 (1.38) &   1.45$^a$  &       6.89 &  2 \\
  88& 222&0+ &0+&2.10 &    1.81 &    1.58 &1.72&    1.39 (1.50)&    1.58$^a$  &    6.68 &  0 \\
  88& 223& 3/2+&5/2+&5.36  &   5.50   &  5.68  & 5.35&  5.17 (5.22) &   5.99$^a$   &     5.98 &  2 \\
  88& 224&0+ &0+&6.02  &   5.71  &   5.53  &5.74&   5.34 (5.41) &   5.50$^a$   &     5.79 &  0 \\
  89& 225&(3/2-) &5/2-&5.99  &   5.95  &   6.05 &  5.52&  5.66 (5.72) &   5.94$^a$  &      5.94 &  2 \\
  88& 226&0+ &0+&11.19  &  10.87  &  10.70 &10.98&   10.56(10.60)&    10.70$^a $ &  4.87 &  0 \\
  90& 228&0+ &0+&8.27   &  7.96  &   7.86 &8.07&    7.71 (7.75)&   7.78$^a$  &     5.53 &  0 \\
  91& 231&3/2- &3/2-&10.93 &   10.89 &   11.34 &11.57&   10.69 (10.71)&   12.01$^a$  &  5.15 &  0 \\
  90& 230&0+ &0+&12.81  &  12.48 &   12.40  &12.75&  12.26 (12.28)&   12.38$^a$    &  4.78 &  0 \\
  92& 232&0+ &0+&9.82  &   9.50 &    9.49&9.69    & 9.34 (9.36)  &  9.34$^a$  &     5.42  & 0 \\
  92& 233&5/2+ &5/2+&12.90   & 13.06   & 13.63 & 12.87&  12.92 (12.91) & 12.70$^a$   & 4.92 &  0 \\
  92& 234&0+ &0+&13.29  &  12.95 &   12.97 &13.21&   12.84 (12.83)&   12.89$^a$  &  4.86 &  0 \\
  94& 236&0+ &0+&8.27   & 7.95  &  8.04  & 8.01& 7.87 (7.87)&  7.95$^a$   &  5.87 &  0 \\
  93&237&5/2+ &3/2-&13.18   & 13.13  &  13.62  &13.59&  13.09 (13.05) &  13.83$^a$   &  4.96 &  1 \\
  94&238&0+ &0+&9.69   &  9.37   &  9.49  &9.54 &  9.30 (9.31)  &  9.44$^a$   &    5.60 &  0 \\
  95&241&5/2- &5/2+&9.98   &  9.94 &   10.54  &10.08&  9.95 (9.92)&   10.13$^a$   &     5.64 &  1 \\
  96&242&0+ &0+&7.34   &  7.02 &    7.22  & 7.11&  7.01 (7.02)&    7.15$^a$  &     6.22 &  0 \\
  90&226&0+ &0+&3.81   &  3.52   &  3.38 &3.49&    3.19 (3.27)  &  3.27$^b$   &     6.46 &  0 \\
  90&232&0+ &0+&18.16  &  17.80  &  17.71&18.18  &  17.63 (17.62) &  17.65$^b$  &  4.08 &  0 \\
  92&230&0+ &0+&6.77 &    6.46 &    6.42 & 6.54&   6.26 (6.30)&     6.25$^b$    &    6.00 &  0 \\
  92&235&7/2- &5/2+&14.54 &   14.69 &   15.21 &15.94&   14.61 (14.57) &  16.35$^b$   & 4.69 &  1 \\
  92&236&0+ &0+&15.24  &  14.90 &   14.94&15.23&    14.80 (14.78)&   14.87$^b$     &  4.58  & 0 \\
  94&240&0+ &0+&11.65   & 11.32  &  11.47 &13.39&   11.28 (11.27)&   11.32$^b$  &  5.26 &  0 \\
  54&112&0+ &0+&3.74 &    3.49   &  -.32 &  &  1.71 (1.90)  &  2.51   &   3.33 &  0 \\
  70&158&0+ &0+&6.94   &  6.66  &   4.66  &&   5.45 (5.62) &   6.63   &    4.18 &  0 \\
  72&160&0+ &0+&3.79   &  3.52  &   1.68  &&   2.37 (2.54) &   2.75    &  4.91 &  0 \\
  74&164&0+ &0+&2.89   &  2.62   &   .93   & & 1.54 (1.71)   & 2.36   &  5.28 &  0 \\
  80&178&0+ &0+&-.06   &  -.33  &  -1.58  & & -1.21 (-1.04) &   -.44   &   6.58 &  0 \\
  85&215&9/2- &9/2-&-3.98  &  -4.01  &  -3.92&-4.48 &   -4.65 (-4.47)&   -4.00 &     8.18  & 0 \\
  86&215&9/2+ &9/2+&-5.39  &  -5.28  &  -5.06  &-5.92&  -5.92 (-5.72) &  -5.64  &     8.85 &  0 \\
  86&216&0+ &0+&-3.68  &  -3.94  &  -4.37 &-4.21&   -4.52 (-4.35) &  -4.35  &     8.20 &  0 \\
  86&217&9/2+ &9/2+&-2.80 &   -2.68   & -2.39  &  -3.31&-3.27 (-3.12) &  -3.27   &   7.89 &  0 \\
  86&218&0+ &0+&-.84   & -1.11   & -1.46 &-1.26  & -1.65 (-1.50) &  -1.46 &     7.27 &  0 \\
  86&219&5/2+ &9/2+& .52    &  .64   &   .74    &.16&  .17 (.29)   &  .60   &    6.95 &  2 \\
  86&220&0+ &0+&2.35&  2.07   &  1.78    &2.02& 1.58 (1.69)   & 1.75   &     6.41  & 0 \\
  86&222&0+ &0+&6.04  &   5.74 &    5.49  &&   5.29 (5.39) &   5.52  &     5.60  & 0 \\
  87&216&(1-) &(1-,9-)&-5.87   & -5.37 &   -5.49  &-6.44&  -6.01 (-5.83)&    -6.15 &      9.18 &  0 \\
  87&217&9/2- &9/2-&-4.08  &  -4.12 &   -3.98&-4.62   & -4.69 (-4.52)  & -4.66  &     8.47 &  0 \\
  87&218&(1-) &1-,9-&-2.83  &  -2.30   & -2.35&-3.30  &  -2.88 (-2.73) &  -3.00   &    8.02 &  0 \\
  87&219&9/2- &9/2-&-1.09  &  -1.13   &  -.92&-1.52   & -1.65 (-1.52)&   -1.70 &     7.46  & 0 \\
  87&220&1+ &(1-)&1.27   &  1.84  &   1.75 &0.83&    1.34 (1.46)&    1.44    &   6.81  & 1 \\
  88&217&(9/2+) &(9/2+)&-5.50   & -5.39  & -5.14& -6.08 &  -5.97 (-5.80)  & -5.80   &   9.16 &  0 \\
  88&218&0+ &0+&-3.96  &  -4.22  & -4.61& -4.49 &   -4.74 (-4.58) &  -4.59  &    8.55 &  0 \\
  88&219&(7/2+) &9/2+&-2.56   & -2.43  &  -2.34&-2.14  &  -2.91 (-2.77) &  -2.00  &    8.13  & 2 \\
  88&220&0+ &0+&-1.18  &  -1.45 &   -1.75 & -1.63&  -1.92 (-1.78)&   -1.60  &    7.60  & 0 \\
  89&217&9/2- &9/2-&-6.74  &  -6.77  &  -6.71  &-7.31&  -7.34 (-7.15)&   -7.16   &   9.84 &  0 \\
  89&218&(1-) &(1-,9-)&-5.71  &  -5.20 &   -5.29 &-6.32&   -5.79 (-5.62)&   -5.95  &     9.38 &  0 \\
  89&219&9/2- &9/2-&-4.37   & -4.40  &  -4.23 &-4.91 &  -4.91 (-4.76)  & -4.93   &   8.83  & 0 \\
  89&220& &(1-)&-3.08  &  -2.55 &   -2.56 &-3.55&   -3.08 (-2.94) &  -1.58   &    8.35  & 0 \\
  89&221&(3/2-) &9/2-&-.63    & -.66  &  -1.20 &  -1.84&  -.92 (-.84)  & -1.28   &   7.79  & 4 \\
  89&222&(1-) &(1-)&.77   &  1.34 &    1.37 &.36&     .87 (0.98) &    .70  &    7.14  & 0 \\
  89&223&(5/2-) &9/2-&2.31  &   2.28  &   2.35&1.78    & 1.94 (2.02)&     2.10 &       6.79 &  2 \\
  89&224&0- &1+&4.08    & 4.68  &   4.65  &3.99&   4.30 (4.38)&    5.06 &     6.32 &  1 \\
  89&226&(1) &2-&8.06   &  8.68   &  8.64  &8.15&   8.36 (8.40) &   9.24  &    5.50 &  1 \\
  89&227&3/2- &(3/2-)&10.57   & 10.53  &  10.89 &10.42&   10.24 (10.27)&   10.70      & 5.05 &  0 \\
  90&217&(9/2+) &1/2-&-4.33  &  -4.22  &  -5.12&  &  -4.44 (-4.36) &  -3.60   &    9.43 &  5 \\
  90&219&&(9/2+)&-5.73   & -5.61  &  -5.33 &-6.29&   -6.14 (-5.96)&   -5.98 &     9.52 &  0 \\
  90&220&0+&0+&-4.37&    -4.63  &  -4.99  &-4.89&  -5.10 (-4.94)&   -5.01 &     8.96 &  0 \\
  90&221&(7/2+) &(9/2+)&-3.31  &  -3.19   & -3.06 &-3.08 &  -3.60 (-3.48)  & -2.77     & 8.64  & 2 \\
  90&222&0+ &0+&-2.10 &   -2.37  &  -2.65& -2.58&   -2.79 (-2.66)&   -2.55   &  8.13  & 0 \\
  90&223&(5/2+) &(7/2+)&-.14     &-.01   &   .18 &  -.25&  -.37 (-.26) &   -.22  &      7.58  & 2 \\
  90&224&0+ &0+&.54   &   .26  &    .05   &  .17&-.12 (-.01)   &  .02   &   7.31 &  0 \\
  90&225&(3/2+) &5/2+&2.20   &  2.32   &  2.57 &2.95 &   2.03 (2.10) &   2.76  &  6.92&   2 \\
  90&229&5/2+ &1/2+&10.56  &  10.72 &   10.98&11.66    &10.54 (10.53)  & 11.36  & 5.17  & 2 \\
  91&215& &&-1.70 & -1.73  &  -1.62&  &  -2.16 (-2.05)&   -1.85  &  8.17  & 0 \\
  91&217& &&-2.71   & -2.74 &   -2.60&-3.21   & -3.18 (-3.06)   &-2.31   &  8.50  & 0 \\
  91&218& &&-6.02  &  -5.51 &   -5.62&-6.65&    -6.05 (-5.89)&    -3.92   & 9.80  & 0 \\
  91&222& &(1-)&-3.36  &  -2.83   & -2.82  &  &-3.31 (-3.17) &   -2.54   & 8.70  & 0 \\
  91&223& &9/2-&-2.40   & -2.43  &  -2.16 &   -2.85 &-2.84 (-2.72)  & -2.19   &  8.35  & 0 \\
  91&224& &&-.16     & .39   &   .46  &-.74&   -.01 (.08)   & -.02    & 7.63 &  0 \\
  91&225& &(3/2-)&.61    &  .57    &  .92   &.25&   .22 (.32)  &   .23  &    7.40   &0 \\
  91&226& &(1-)&2.09   &  2.67    & 2.77  &1.69&   2.31 (2.39)&    2.16   &  6.99 &  0 \\
  91&227&(5/2-) &(5/2-)&3.69   &  3.65 &    4.05&3.36  &   3.35 (3.42) &   3.43    & 6.59 &  0 \\
  91&229&(5/2+) &(3/2-)&7.20 &    7.16  &   7.51&7.69     &6.93 (6.96)  & 7.43    &  5.84 &  1 \\
  92&226&0+ &0+&.40     & .12   &  -.05   &-.45&  -.19 (-.09)   & -.30   &   7.57  & 0 \\
  92&228&0+ &0+&3.19  &   2.90  &   2.80&2.86     &2.64 (2.71)    &2.76    &  6.81&   0 \\
  92&229&(3/2+) &(3/2+)&4.56    & 4.70  &   5.26 &4.27&    4.45 (4.51)&     4.24&    6.48&   0 \\
  92&231&(5/2-) &(1/2+)&9.53   &  9.68  &   9.77  &8.89&   9.64 (9.62) &   9.82   &  5.56 & 3 \\
  93&229& &&2.81    & 2.77&     3.21  &2.43&   2.54 (2.59) &   2.68  &   7.01 &  0 \\
  93&231&(5/2) &(5/2-)&5.47    & 5.43   &  5.91&5.17     &5.24 (5.27)   & 5.17  &    6.37  & 0 \\
  93&235&5/2+ &3/2-&11.68    &11.64  &  12.09 &11.98&   11.56 (11.54)&   12.12 &    5.20 &  1 \\
  94&232&0+ &0+&4.36  &   4.06    & 4.05   &4.04&  3.91 (3.95)   & 4.01  &    6.72  & 0 \\
  94&233& &(3/2+)&5.66    & 5.80   &  6.43 &5.38&    5.65 (5.68) &   6.02  &    6.42   &0 \\ 
  94&234&0+ &0+&6.09   &  5.79    & 5.83   &5.88&  5.67 (5.70) &   5.72   &    6.32 &  0\\ 
  94&235&(5/2+) &(5/2-)&7.90    & 8.04   &  8.64&7.34    & 7.96 (7.95)   & 7.75  &     5.96 &  1\\ 
  94&237&7/2- &5/2+&8.97 &    9.11    & 9.73&10.37   &  9.05 (9.04) &  10.97   &   5.75 &  1\\ 
  94&239&1/2+ &1/2+,7/2-&11.79 &   11.94  &  12.66 &11.66&   11.89 (11.87)&   11.88   &  5.24 &  0\\ 
  95&238&1+ &(0+)&7.79    & 8.42   &  8.71   &7.58&  8.34 (8.34)  &  9.77   &    6.05 &  0\\ 
  95&239&(5/2-) &5/2+&8.51   &  8.47  &   9.03 &8.47&    8.45 (8.44)&    8.63 &    5.92 &  1\\ 
  95&240&(3-) &(6-)&10.33   & 10.98&    10.49 &10.04&   11.18 (11.07)  & 10.98   &   5.71 &  4\\ 
  95&242&5- &2+&10.59   & 11.23&    11.10&   & 11.35 (11.28)&    11.99&   5.60 &  3\\ 
  95&243&5/2- &5/2+&11.09   & 11.05  &  11.68 &&   11.08 (11.04) &  11.37   & 5.44  & 1\\ 
  96&238&0+ &0+&5.56  &   5.25  &   5.37  &5.67&   5.22 (5.22)  &  4.94 &   6.62&   0\\ 
  96&240&0+ &0+&6.52    & 6.21    & 6.37&6.25     &6.19 (6.18)    &6.37   &  6.40  & 0\\ 
  96&241&1/2+ &1/2+,7/2-&7.50    & 7.65   &  8.44  &8.04&   7.63 (7.62)&    8.45  &   6.19 &  0\\ 
  96&243&5/2+ &1/2+&7.75&     7.88  &   8.49    & 8.87&7.95 (7.93)  &  8.96   &6.18  & 2\\ 
  96&244&0+ &0+&8.87  &   8.54  &   8.79&8.68     &8.58 (8.56)   & 8.76   &  5.91 &  0\\ 
  97&245&3/2- &5/2-&6.82   &  6.78  &   7.31&     &6.89 (6.86)   & 8.55   &  6.46&   2\\ 
  97&247&(3/2-) &5/2-&9.48     &9.43  &  10.15& &    9.53 (9.50)&    10.64 &  5.89&   1\\ 
  98&242&0+ &0+&2.68   &  2.39   &  2.55 &2.34   & 2.39 (2.41)  &  2.32  &  7.52  & 0\\ 
  98&244&0+ &0+&3.32    & 3.02   &  3.24   &2.99&  3.04 (3.06) &   3.07  &    7.34&   0\\ 
  98&250&0+ &0+&8.51  &   8.19  &   8.55&     &8.32 (8.27)    &8.62      &6.14  & 0\\ 
  98&251&1/2+ &9/2-&9.43    & 9.59    & 9.42 & 10.32&   10.09 (9.92)&   10.45&      6.18 &  5\\ 
  98&252&0+ &0+&8.07    & 7.75   &  8.15 &7.88   & 7.88 (7.84)   & 7.93    & 6.22  & 0\\ 
  98&253&(7/2+) &(1/2+)&9.31  &   9.47 &    9.72&  &   9.84 (9.71)  &  8.70    & 6.12  & 4\\ 
  98&254&0+ &0+&9.47   &  9.15  &   9.59   &&  9.30 (9.27)  &  9.23    & 5.94  & 0\\ 
  99&251&(3/2-) &(3/2-)&6.68    & 6.64     &7.52 &  &  6.78 (6.74) &   7.38 &  6.61  & 0\\ 
  99&252&(5-) &(6+,1-)&6.06   &  6.68  &   7.18  & &  6.81 (6.78)   & 7.73   & 6.76 &  1\\ 
  99&253&7/2+ &7/2+&6.01   &  5.97 &    6.88 &   & 6.10 (6.08)&     6.25&  6.75   &0\\ 
  99&255&(7/2+) &(3/2-)&7.85  &   7.82 &    8.31    & &8.11 (8.03)    &7.63  &   6.44  & 3\\ 
 100&251&(9/2-) &(7/2+)&3.73   &  3.88   &  4.77  & &  4.01 (3.99)&     6.03&  7.43  & 1\\ 
 100&252&0+ &0+&4.76    & 4.45 &    4.83 &  &  4.60 (4.58)  &  4.96  &  7.15  & 0\\ 
 100&253&1/2+ &9/2-&5.64    & 5.79    & 5.69&4.84  &   6.31 (6.15)&     6.33&   7.20 &  5\\ 
 100& 254&0+ &0+&4.08   &  3.77   &  4.18 &   & 3.93 (3.91)  &  4.07   &  7.31 &  0\\ 
 100&255&7/2+ &1/2+&5.07    & 5.21   &  5.52 & &   5.60 (5.51) &   4.86   & 7.24&   4\\ 
 100&256&0+ &0+&5.18   &  4.87   &  5.33 &  &  5.04 (5.02)   & 5.07   & 7.03&   0\\ 
 100&257& (9/2+)&(7/2+)&6.06     &6.22 &    7.08 & &   6.46 (6.40) &   6.94& 6.87 &  2\\ 
 102&254&0+ &0+&1.52    & 1.22   &  1.58&1.12   &  1.39 (1.38)   & 1.79   & 8.23&   0\\ 
 102&255&(1/2+) &(9/2-)&1.84    & 1.99&     1.91&1.04  &   2.50 (2.36)&    2.48 &8.45 &  5\\ 
 102&256&0+ &0+&.31     & .02    &  .39   &-.07&   .17 (.19)  &  .52    &8.59 &  0\\ 
 102& 257&(7/2+) &1/2+&1.42    & 1.56   &  1.87&.34  &   1.94 (1.87)  &  1.40 & 8.46  & 4\\ 
 102&259&(9/2+) &7/2+&3.11    & 3.24    & 4.13   &3.29  &3.52 (3.48)  &  3.67 &7.81 &  2\\ 
 106&263& &&-.81    & -.68   &   .39  &  & -.40 (-.41)&     .43 &   9.40 &  0\\ \hline

$\chi^2/F$&&&&96.31&74.86&78.33&224.13&7.84 (4.77)&&& \\ 

\end{tabular} 
$(a)$, $(b)$  Experimental data from references \cite{r13}, \cite{r14} respectively, and rest of the experimental data from reference \cite{r6}.
\end{table}
    
      The decay modes and the experimental values for their half lives from references \cite{r14} and \cite{r17} for heavier clusters have been presented in Table 2. The released energy Q have been calculated using the experimental ground state masses from \cite{r18} . Whenever the experimental ground state masses are not available, the theoretically calculated ground state masses from the latest mass table \cite{r18} have been used. The corresponding results of present calculations using microscopic potentials within the superasymmetric fission model description have also been presented along with the results of the ASAFM(1986), ASAFM(1991) and the liquid drop model (LDM) \cite{r17}. The results of ASAFM(1986) and the  ASAFM(1991) have been recalculated with the Q values listed in the Table 2. The $\chi^2/F$ for Table 2 have been calculated assuming an uniform two percent experimental error for the entire data set, but for those data that represent only the lower limits for the decay half lives, an uniform ten percent error in the measurements have been  assumed. 
                                 
\begin{table}
\caption{Comparison between Measured and Calculated heavier cluster decay Half-Lives}
\begin{tabular}{cccccccccccc}
Parent&Emitted&Parent&Daughter&Emitted&ASAFM(86)& ASAFM(91)&LDM(01)&Present&Expt.& Energy&$l_{min}$ \\ 
 & & & & &  & & &M3Y & &Released & \\ \hline
Z,A&$Z_d$,$A_d$&$J^{\pi}$&$J^{\pi}$&$J^{\pi}$&$log_{10}T(s)$&$log_{10}T(s)$&$log_{10}T(s)$&$log_{10}T(s)$&$log_{10}T(s)$&Q(MeV)& \\ \hline
  87, 221& 6,  14&5/2-&1/2+&0+&   15.14 &   14.51&13.68  &  12.70 &   14.5 &  31.30 &  3 \\
  88, 221& 6,  14&5/2+&1/2-&0+&   13.98  &  14.37 & 12.18 & 12.42 &   13.4  &   32.40 &  3 \\
  88, 222& 6,  14&0+&0+&0+&  12.56 &    11.16 &  10.59  &9.59 &   11.0 &    33.05 &  0 \\
  88, 223& 6,  14&3/2+&9/2+&0+&  15.02 &   15.42 &13.45&   13.50 &   15.2 &  31.85 &  4 \\
  88, 224& 6,  14&0+&0+&0+&   17.39 &   15.95  &16.59 & 14.34 &   15.7& 30.53 &  0  \\
  88, 226& 6,  14&0+&0+&0+&   22.43 &   20.97 &22.51  & 19.37  &  21.2&  28.21 &  0 \\
  89, 225& 6,  14&(3/2-)&9/2-&0+&   18.67&    18.04 & 17.81&  16.43 &   17.2& 30.48 &  4 \\
  90, 226& 8,  18&0+&0+&0+&   18.95  &  18.05 & 18.95 & 16.66 &   $>$16.8& 45.73 &  0 \\
  90, 228& 8,  20&0+&0+&0+&   22.44 &   21.95 & 21.61 & 19.89 &   20.7&  44.72 &  0 \\
  90, 230& 10,  24&0+&0+&0+&   24.86  &  25.27 &25.45  & 23.39 &   24.6&  57.78 &  0 \\
  90, 232& 10,  24&0+&0+&0+&   28.14&    28.55 &  & 26.66 &   $>$29.0&     55.76 &  0 \\
  90, 232& 10,  26&0+&0+&0+&   29.36 &   30.24  &29.72 & 27.70   & $>$29.0&  55.97&  0 \\
  91, 231& 9, 23&3/2-&0+&(3/2+,5/2+)&   24.73  & 25.88 &24.26 &  23.47  &  $>$26.0  &   51.86&   1 \\
  91, 231& 10,  24&3/2-&1/2+&0+&   22.00   & 23.40&21.93 &   21.42 &   22.9 &  60.42&   1 \\
  92, 230& 10,  22&0+&0+&0+&   20.51   & 20.44 &21.40  & 19.48 &   19.6&  61.40&  0 \\
  92, 232& 10,  24&0+&0+&0+&   20.41  &  20.81 & 19.99 & 19.35&   20.4&  62.31&  0 \\
  92, 233& 10,  24&5/2+&9/2+&0+&   23.15 &   24.84 &23.36 &  22.94  &  24.8&     60.51&   2 \\
  92, 233& 10,  25&5/2+&0+&(1/2+,3/2+)&   23.45   & 25.20 &23.15 &  23.06 &   24.8&     60.75&   2 \\
  92, 233& 12,  28&5/2+&1/2-&0+&   24.55 &   26.48  & 25.78& 25.04  &  $>$27.6&    74.25&  3 \\
  92, 234& 10,  24&0+&0+&0+&   25.72 &   26.13 &26.54 &  24.59 &   $>$26.0&    58.84&   0  \\
  92, 234& 10,  26&0+&0+&0+&   26.16 &   27.05 & 25.91 & 24.91 &   $>$26.0&    59.47&   0 \\
  92, 234& 12,  28&0+&0+&0+&   24.56  &  25.03  &25.90 & 24.03  &  25.7&   74.13&   0 \\
  92, 235& 10,  24&7/2-&9/2+&0+&   28.12   & 29.83  &29.40 & 27.85   & 27.4&    57.36&   1 \\
  92, 235& 10, 25&7/2-&0+&(1/2+,3/2+)&   28.39  &  30.10  &29.08 & 27.96   & 27.4&   57.73&   3 \\
  92, 235& 12,  28&7/2-&(9/2+)&0+&   27.31  &  29.31 &29.26 &  27.75 &  $>$28.0 &   72.21&   1 \\
  92, 236& 12,  28&0+&0+&0+&   27.82   & 28.29 &  & 27.24 &   27.6&    71.83&  0 \\
  92, 236& 12,  30&0+&0+&0+&   28.09  &  28.57  &29.28 & 27.21&    27.6&    72.48& 0 \\
  93, 237& 12,  30&5/2+&1/2+&0+&   25.84  &  27.58 &26.56 &  26.10 &   $>$27.6&   74.99&  2 \\
  94, 236& 12,  28&0+&0+&0+&   19.79  &  20.26  & 20.00 & 19.93   & 21.7&   79.67&   0 \\
  94, 238& 12,  28&0+&0+&0+&   24.81 &   25.29   &26.34 &24.72 &  25.7  & 75.93&   0 \\
  94, 238& 12,  30&0+&0+&0+&   24.42   & 24.91 & 24.83 & 24.13 &   25.7  &  77.00&   0 \\
  94, 238& 14,  32&0+&0+&0+&   23.69    &24.23  &25.73 & 24.52 &   25.3&   91.21&   0 \\
  94, 240& 14,  34&0+&0+&0+&   24.64  &  25.19   & 26.08& 25.34 &   $>$25.5& 91.05&   0 \\
  95, 241& 14,  34&5/2-&1/2+&0+&   22.47 &   24.46 &23.32 &  24.45   & $>$25.3&  93.94&   3 \\
  96, 242& 14,  34&0+&0+&0+&   20.75 &   21.31 & 21.11 &22.14 &   23.2&   96.53&   0 \\ 
  90, 226&  6,  14&0+&0+&0+&  19.26 &   17.79  &18.79 & 16.36 &   $>$15.3$^c$   &30.55&   0 \\
  92, 230& 10,  24&0+&0+&0+&   22.03 &   22.43  &21.97 & 20.85 &  $>$ 18.2$^c$  & 61.36&   0 \\
  92, 232& 12,  28&0+&0+&0+&   24.46  &  24.93 &25.74 &  23.85   & $>$22.7$^c$   &  74.33&   0 \\
  92, 236& 10,  24&0+&0+&0+&   30.51&    30.93  &32.18 & 29.31   &$>$ 26.0$^c$   & 55.96&   0 \\
  92, 236& 10,  26&0+&0+&0+&  30.76   & 31.65 &31.48 &  29.42  & $>$26.0$^c$& 56.75&   0 \\ \hline

 $\chi^2/F$&&&&&6.61&3.60&6.30&7.05&&& \\ 

\end{tabular} 
$(c)$ Experimental data from reference \cite{r17} and rest of the experimental data from reference \cite{r14}.
\end{table}
     
      The $\chi^2/F$ can be brought down from 7.84 to a minimum of 4.42 for $\alpha$ and from 7.05 to 2.21 for heavier clusters by adjusting the depths of the microscopic nuclear potentials using normalization constants of 0.979 and 0.870 respectively. However, for the present calculations depths of the nuclear potentials obtained by double folding the M3Y effective interaction have not been adjusted.   

      The half lives for cluster-radioactivity have been analyzed with microscopic nuclear potentials which are based on profound theoretical basis. It is worthwhile to mention that these reasonably exhaustive calculations using realistic microscopic nuclear interaction potentials have been performed without adjusting the depth of the nuclear potentials using any renormalization or adjusting any other parameters. Considering the fact that the $\alpha$ particles can be detected rather easily under favourable conditions such as high efficiency, low background and good energy resolution as compared to the heavier clusters where experimental uncertainties are more, the results of the present calculations of SAFM using micoscopic potentials are in excellent agreement over a wide range of experimental data spanning about thirtyfive orders of magnitude. Such calculations can be used to provide reasonable estimates for the lifetimes of nuclear disintegration processes into two composite nuclear fragments for the entire domain of exotic nuclei.

\end{document}